\documentclass[referee]{aa}
\usepackage{graphicx}

\begin{document}

\def\pdot {\dot P}
\def\Omdot {\dot \Omega}
\def\ltsima{$\; \buildrel < \over \sim \;$}
\def\lsim{\lower.5ex\hbox{\ltsima}}
\def\gtrsim{$\; \buildrel > \over \sim \;$}
\def\gsim{\lower.5ex\hbox{\gtsima}}
\def\msole{~M_{\odot}}
\def\mdot {\dot M}
\def\ee {1E~1207.4--5209~}
\def\cha {\textit{Chandra~}}
\def\snr {G 296.5+10.0~}
\def\xmm  {\textit{XMM-Newton~}}

\title{XMM-Newton and VLT observations of the isolated neutron star  \ee\thanks{Based on observations with \xmm,
 an ESA science mission with instruments and contributions directly  funded by ESA member states and the USA (NASA),
 and on observations collected at the European Southern Observatory, Paranal, Chile, under proposal 69.D-0528(A)}}
\author{A. De Luca\inst{1,2} \and S.Mereghetti\inst{1} \and P.A.Caraveo\inst{1} \and M.Moroni\inst{1,3}
\and R.P. Mignani\inst{4} \and G.F. Bignami\inst{3,5}}
\institute{Istituto di Astrofisica Spaziale e Fisica Cosmica,
Sezione di Milano  ''G.Occhialini'' - CNR
v.Bassini 15, I-20133 Milano, Italy
\and Universit\`a di Milano Bicocca, Dipartimento di Fisica, P.za della Scienza 3, 20126 Milano, Italy
\and Universit\`a degli Studi di Pavia, Dipartimento di Fisica Nucleare e Teorica, Via Bassi 6, 27100 Pavia, Italy
\and European Southern Observatory, Karl Schwarzschild Strasse 2, D-85740, Garching, Germany
\and Centre d'Etude Spatiale des Rayonnements, CNRS-UPS, 9, Avenue du Colonel Roche,  31028 Toulouse Cedex 4, France}

\offprints{A. De Luca,
\email{deluca@mi.iasf.cnr.it}}
   \date{Received ...; accepted ...}

\abstract{

In August 2002, XMM-Newton devoted two full orbits to the observation of \ee,
 making this isolated neutron star the most deeply scrutinized galactic target
 of the mission.
Thanks to the high throughput of the EPIC instrument, $\sim$
360,000 photons were collected from the source, allowing for a
very sensitive study of the temporal and spectral behaviour of
this object. The spectral data, both time-averaged and
phase-resolved, yield one compelling interpretation of the
observed features: cyclotron absorption from one fundamental
($\sim 0.7$ keV) and three harmonics, at $\sim1.4$, $\sim2.1$ and
$\sim2.8$ keV. Possible physical consequences are discussed, also
on the basis of the obvious phase variations of the features'
shapes and depths. We also present deep VLT optical data which we
have used to search for a counterpart, with negative results down
to $V\sim27$.

}
\titlerunning{XMM and VLT observations of \ee}
   \maketitle

\section{Introduction}

The X-ray source \ee attracted much interest since its early discovery
(Helfand \& Becker 1984) as  a bright unresolved source located at the
geometrical  center of  the  shell-like radio/X-ray/optical  supernova
remnant \snr\ (Roger et al. 1988).
X--ray observations with the Einstein (Helfand \& Becker 1984), EXOSAT
(Kellet et al. 1987), ROSAT (Mereghetti et al. 1996) and ASCA (Vasisht
et    al.     1997)   satellites    showed    a    steady   flux    of
$\sim$2$\times$10$^{-12}$   erg   cm$^{-2}$   s$^{-1}$   (0.3-3   keV)
characterized by a thermal spectrum.
These observations,  coupled with the  lack of an  optical counterpart
down to $V \sim 25$ (Bignami  et al. 1992; Mereghetti
et al. 1996)
strongly suggested a neutron star nature for \ee.
This  hypothesis was  later confirmed  by the  \cha detection  of fast
X--ray pulsations  with period  $P$=0.424 s (Zavlin  et al.  2000) and
period derivative $\pdot$  = (2.0$^{+1.1}_{-1.3}$) $\times$ 10$^{-14}$
s s$^{-1}$ (Pavlov et al. 2002).

The \cha data  also unveiled the presence  of broad absorption
features at  energies  of $\sim$0.7  and $\sim$1.4 keV (Sanwal et
al. 2002). The  existence  of these features  was soonafter
confirmed by  an \xmm observation, which showed  that the  depths
and  profiles of  the two lines vary significantly with  the
rotational  phase  of the  pulsar (Mereghetti et al. 2002a,
hereafter Paper I).

Although observationally firmly  established, the nature  of these
lines could not be unambiguously  identified.   They  were
attributed either  to  HeII transitions         in        a
``Magnetar''-like        field
B$\sim$(1.4--1.7)$\times$10$^{14}$  G  (Sanwal et al. 2002) or
to atomic transitions  in heavier elements (e.g. He-like Oxigen
or Neon) in the atmosphere of a  neutron star with a more
conventional magnetic field (Hailey \& Mori 2002).  An alternative
explanation of the lines as cyclotron features was considered to be 
unlikely by Sanwal et al. (2002). This view was criticized by 
 Xu et al. (2003), who interpreted the
lines as electron cyclotron resonance features originating near
the neutron star surface.

The breakthrough came  with a longer \xmm observation  which,
besides confirming  the  two phase-dependent  absorption lines at
0.7 and 1.4 keV, showed a statistically significant third line at
$\sim$2.1 keV, as  well as  a hint  for a  possible fourth
feature at  2.8  keV  (Bignami et  al. 2003, hereafter  Paper
II).  The nearly 1:2:3:4 ratio  of the line centroids,  as well
as  the phase variation, naturally following the pulsar B-field
rotation, strongly suggest that such lines are due to cyclotron
absorption  processes in a magnetic field of
$\sim$8$\times10^{10}$ G or $\sim$1.6$\times10^{14}$ G,
respectively in the case of electrons or protons features.

Thus, among known isolated neutron stars, \ee stands out as the
only one which clearly exhibits X--ray absorption lines.

Here we present a comprehensive analysis of the data collected
during the long \xmm observation, already shortly discussed in
Paper II, as well as new optical images of the field, the deepest
available so far, performed with the VLT.

\section{XMM-Newton Data reduction and analysis}

\label{xmmdata}
The \xmm observation of \ee started on August 4, 2002
and lasted two full  orbits yielding
 two uninterrupted time intervals of $\sim$36 hours each.

The data reported here were obtained with the
European Photon Imaging Camera (EPIC) instrument, which consists
of two MOS CCD detectors (Turner et al. 2001) and a pn
CCD instrument (Str\"{u}der et al. 2001), for a total collecting
area \gtrsim 2500 cm$^2$ at 1.5 keV. The mirror system offers an
on-axis point spread function of 4-5$''$ FWHM and a field of view of 30$'$
diameter.

While the two MOS   cameras were operated in ``full frame'' mode,
the pn camera was operated in ``small window'' mode to allow for the accurate
timing of the source photons (6 ms resolution).
All cameras used the thin filter.
The data were processed with the latest release of the XMM Newton Science Analysis Software
(SAS version 5.4.1). After screening the data to remove time intervals
with high particle background
and correcting for the dead time, we obtain a net exposure time of 139.0
ksec for the pn
camera and 197.5 and 197.3 ks for the MOS1 and MOS2, respectively.

\begin{figure}
\centering
  \includegraphics[width=\textwidth]{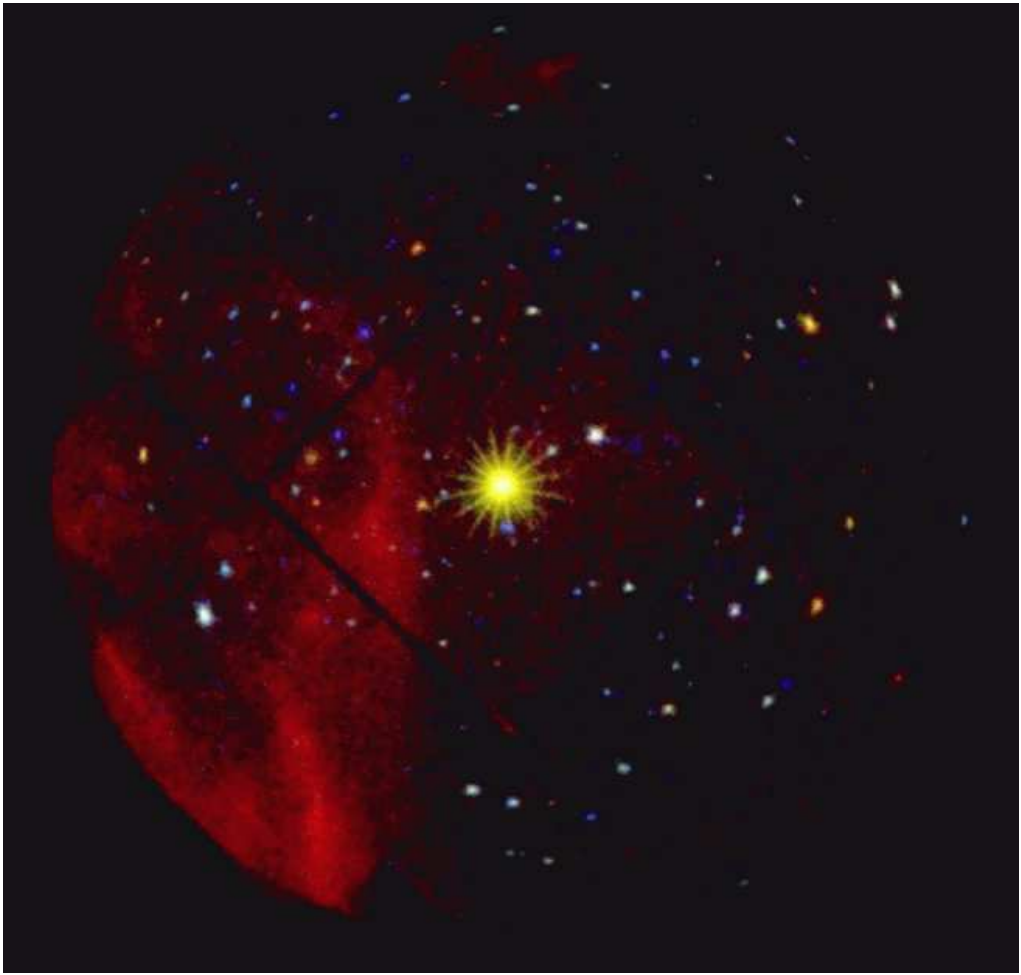}
  \caption{ The XMM-Newton view of the field of \ee.
  Data from the MOS1 and MOS2 camera have been merged to produce the image. The exposure time is of $\sim$200 ksec per camera. The target is the bright
  source close to the center of the field. Many ($\sim$200) serendipitous sources
and the bright emission from parts of the shell of the host supernova remnant G296.5+10.0
are clearly visible. The colors are energy-coded, red represents  photons
in the 0.3-0.6 keV band, green and blue correspond to the 0.6-1.5 keV and 1.5-8 keV bands
respectively}
\label{mosfield}
\end{figure}

The source \ee is clearly detected in the three EPIC cameras.
In the energy range 0.2-5 keV  it yields
net count rates of 1.466$\pm$0.003 cts s$^{-1}$, 0.380$\pm$0.002  cts s$^{-1}$ and 0.389$\pm$0.002 cts s$^{-1}$
 in the pn, MOS1, and MOS2 respectively.

The field of \ee as seen by EPIC is shown in Fig.~\ref{mosfield}. Data from the MOS1
and MOS2 cameras have been merged to produce the image, which has energy-coded
colors (see caption).

In the MOS field of view a large ($\sim 200$) number of serendipitous
sources
are detected. Bright  emission from parts of the surrounding
supernova remnant G296.5+10.0 is also visible.
No extended emission is seen around \ee or in its immediate vicinity (within $\sim$2
arcmin). The radial intensity profile of the target is fully consistent with
the instrumental point spread function (Ghizzardi 2002).

To derive the sky coordinates of \ee we computed the boresight
correction to be applied to the default EPIC astrometry.
This was done independently for the MOS1 and MOS2
cameras; the pn data were not used since the small field of view (4$\times$4
arcmin) prevented the detection of a suitable number of serendipitous sources.

The positions of the $\sim$200 serendipitous sources detected in
the MOS field of view were correlated with the Guide Star Catalog
II
(GSC-II\footnote{http://www-gsss.stsci.edu/gsc/gsc2/GSC2home.htm}).
After rejecting ambiguous matches, the optical brightness and the
X-ray spectrum were used as criteria to identify the X-ray
sources having stellar counterparts. This yielded 6 good
reference sources which were used to correct the EPIC astrometry.
The rms error between the refined X-ray and GSC-II positions was
found to be of $\sim$1$''$ per coordinate, entirely consistent
with the expected $\leq$1.5$''$ internal EPIC astrometric
accuracy  (Kirsch 2003). The resulting MOS1 position of \ee is
$\alpha_{J2000}=12h10m00.91s$,
$\delta_{J2000}=-52^{\circ}26'28.8''$ with an overall error
radius of 1.5$''$, including the quoted residual uncertainty as
well as the absolute intrinsic accuracy (0.35$''$ per coordinate)
of the GSC-II. The MOS2 position is $\alpha_{J2000}=12h10m00.84s$,
$\delta_{J2000}=-52^{\circ}26'27.6"$, with an uncertainty of
1.5$''$, similar to the MOS1 case. The two positions are fully
consistent, with a difference well within the expected accuracy
of the relative astrometry between the MOS cameras.

In order to obtain an independent measurement on the position
of \ee, we have retrieved from the Chandra Data Archive a public
dataset relative to a recent (2003/06/15) ACIS Timed Exposure
mode observation (20 ksec) of the target.

Following the Chandra X-ray Center threads to improve the absolute
astrometry of the standard pipeline-processed archived
data\footnote{http://cxc.harvard.edu/ciao/threads/arcsec\_correction/index.html},
 we used the Aspect
Calculator\footnote{http://cxc.harvard.edu/cal/ASPECT/fix\_offset/fix\_offset.cgi}
to verify that the selected observation was not affected by any
known aspect offset.

Then, as in the case of the EPIC data, we used the positions of
the serendipitous sources in the field to refine the astrometry.
We correlated the ACIS and EPIC source positions in order to
reject any spurious detection in the lower statistic Chandra
observation. Considering the region within $\sim10$ arcmin from
the target, we selected 8 secure coincidences. Their coordinates
were cross-correlated with the GSC-II catalog. Only two sources
were found to have a match within 3$''$. The boresight correction
was found to be of $\sim0.8''$, in agreement with the expected
absolute aspect
accuracy\footnote{http://asc.harvard.edu/cal/ASPECT/celmon/},
with an uncertainty of $\sim0.35''$. The best Chandra/ACIS
position of \ee is $\alpha_{J2000}=12h10m00.826s$,
$\delta_{J2000}=-52^{\circ}26'28.43''$ with an uncertainty of
0.6$''$. The more accurate Chandra position lies inside the
intersection of the MOS1 and MOS2 error circles. This gives us
confidence about the correctness of our analysis and the absence
of systematics.

\begin{figure}
\centering
  \includegraphics[angle=-90,width=\textwidth]{1e1207_p.ps}
  \caption{Period history of \ee . Circles are the \cha measurements,
crosses
  \xmm      }
  \label{pdot}
\end{figure}

\begin{figure}
\centering
  \includegraphics[angle=-90,width=\textwidth]{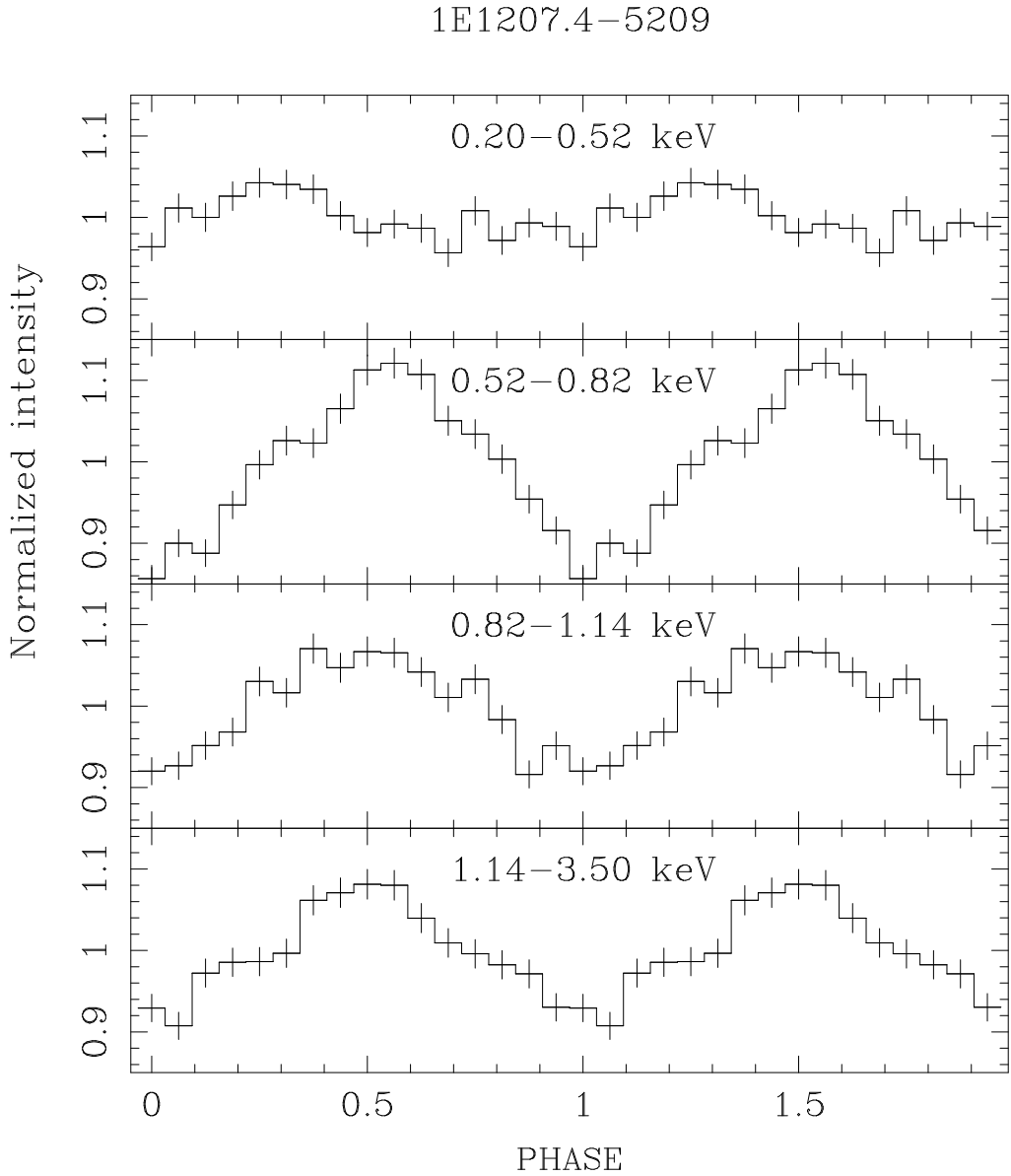}
  \caption{Folded light curve of \ee in four energy ranges.}
\label{lcs}
\end{figure}

\subsection{Timing Analysis}
\label{timing} We based our timing analysis on the pn counts
extracted from a circular region of 43$''$ radius centered on the
source position and with energy in the 0.2 - 3.5 keV interval.
After converting the times of arrival to the  Solar System
Barycenter, we searched the period range from 424.12 to 424.14 ms
using both a folding algorithm with 8 phase bins and the
Rayleigh  test. Both methods yielded a highly significant
detection at P = 424.13076$\pm$0.00002 ms. As in Paper I, the
best period value and its uncertainty were determined following
Leahy (1987) and verified through simulations.

Comparing the new period measurement of  \ee with that obtained with
\cha in January 2000 (Pavlov et al. 2002) we obtain a period
derivative $\pdot$=(1.4$\pm$0.3)$\times$10$^{-14}$ s s$^{-1}$.
This is consistent with the value given in Paper I, but it has a smaller
error.

We note that the $\pdot$ value rests totally on the first \cha
period measurement. Using only the 3 most recent values, the period
derivative is unconstrained, as is clearly seen in Fig.~\ref{pdot}.

To study the energy dependence of the pulse profile, we divided the data in
four channels with approximately 52,000 counts each:
0.2-0.52 keV, 0.52-0.82 keV, 0.82-1.14 keV and 1.14-3.5 keV.
We verified that an independent analysis of each of these channels with the
method
described above would have resulted in a statistically
significant detection of the pulsation.
The  pulse profiles in the different energy ranges (Fig.~\ref{lcs})   show a broad,
nearly
sinusoidal shape, with a  larger pulsed fraction
in  the 0.52-0.82 keV band.

The pulsed fraction in the four energy ranges, defined as the ratio between the number of counts
above the DC level and the total number of counts, is of
(4.0$\pm$0.6)\%,
(13.8$\pm$0.6)\%,
(8.3$\pm$0.6)\% and
(9.5$\pm$0.6)\%
(from the lowest to the highest energy range).

Comparing now the shapes of the light curves, we note that all but the
very soft one show the peak at the same phase.
A phase shift of
nearly 90$^{\circ}$ between the profile in the lowest energy range ($<$0.52
keV) and those at higher energies is apparent.
Indeed, a fit to the light curves with a {\em sin} function
yields best fit phases of
 0.023$\pm$0.026,
0.294$\pm$0.006,
0.245$\pm$0.010,   and
0.255$\pm$0.010 from the lowest to the highest energy range.

We reanalyzed the data of the December 2001 observation to search for the same effect,
but the lower statistics hampers a firm  conclusion. In fact,
 using the $\sim$6600 counts
collected in that observation below 0.5 keV ,the pulsations are only marginally detectable.
Finally we note that the new observation confirms that no
phase shift occurs at $\sim$1 keV, as reported by Pavlov et al. (2002) using
 \cha data.

\subsection{Spectral Analysis}
\label{spec} To perform the spectral analysis we used events
extracted from a 43" radius circle centered on the source,
selecting PATTERN in the range 0$\div$4 for the pn and 0$\div$12
for the MOS. 
 As a consistency check, we verified {\em a posteriori} that the 
results of our analysis do not change when using only PATTERN 0
 (i.e. single pixel) events.
Particular care was devoted to the selection of
the background regions. In the small pn field of view we excised
the source with a circle  of 75$''$ radius and we rejected the
area possibly contaminated by out-of-time events, or too near to
the CCD edges. In the MOS cameras we selected a region within the
central CCD, excluding both  point sources and the diffuse
emission from the supernova remnant, avoiding the CCD edges,
contaminated by Si-K internal fluorescence emission. The spectra
were rebinned in order to have at least 40 counts per channel;
the pn spectrum was rebinned in order to oversample the
instrumental energy resolution by a factor of 3. Ad hoc response
matrices and ancillary files were generated using the SAS tasks
{\em rmfgen} and {\em arfgen}. The spectral analysis was performed
using XSPEC v11.2 in the energy range 0.3-4 keV. Energies lower
than 0.3 keV were discarded both in the MOS, owing to calibration
uncertainties, and in the pn, owing to a strong (possibly
electronic) feature in the background at $\sim$0.22 keV. Beyond 4
keV the source is only marginally detected.

The large number of photons collected in our observation
makes statistical errors very small.
Particular care must be devoted to the systematic uncertainties.
The internal calibration accuracy of each EPIC detector for
on-axis sources is better than 5\% (Kirsch 2003). This yields good quality
fits (i.e. $\leq$5\% residuals at the instrumental edges)
for each camera (see Fig.~\ref{3c273}), but does not ensure the correctness
of absolute flux measurements. Indeed, while the cross-calibration between
the MOS1 and MOS2 cameras agrees within 5\%, differences
up to 10\% are found between the pn and the MOS, the
MOS flux being smaller than the pn one below $\sim$1.5 keV
and higher above this energy.
These effects are probably due to remaining
uncertainties in the vignetting and CCD quantum efficiency
functions (Kirsch 2003).

We therefore adopted the following strategy
for the phase-integrated spectra. As a first step,
the computation of the best fitting model was performed separately
for each instrument, accounting only for statistical errors.
As a second step, to compute reasonable confidence intervals on the
measured physical parameters of the target, we added an extra 5\% systematic
error to each spectral channel.

A different approach was used for the phase-resolved spectral analysis.
For this study only the pn data can be used, owing to the MOS slow readout mode.
Our aim is the description of the relative spectral variations
as a function of the pulse phase,
which are correctly characterized even when the
systematic uncertainties are not included.
Such systematics affect all of the spectra in the same direction (since
they are taken with the same instrument) and are not a matter of concern
when relative differences are studied. Therefore, we do not include systematics
in the evaluation of the confidence intervals for the phase resolved
spectral parameters to give a correct description of their {\em relative} variation,
warning the reader that the quoted errors are possibly underestimated when
the {\em absolute} values of the parameters are considered.

\subsubsection{Phase integrated spectroscopy}

As a first step, we addressed the  phase integrated spectrum with respect
 to our previous analysis (Paper II).
The improved understanding of the instruments (concerning, e.g.,
the Quantum Efficiency, the Charge Transfer Inefficiency, the redistribution
function)
implemented in the most recent SAS
release yields significant differences in the low energy portion of the
source spectrum, below the O edge ($\sim$0.55 keV), with respect to our previous
analysis. As a consequence, the spectral parameters reported here supersede
the results of Paper II.

The spectrum of \ee is very complex. Single component (blackbody,
power law, ...) or double component (blackbody+power law, blackbody+blackbody...)
continuum models alone are totally inadequate. A satisfactory fit
requires a model including broad absorption features.
 In the following we will discuss separately the continuum
and the line components, starting from the pn, which collected $\sim$
194,000 photons in the 0.3-4 keV band.

\begin{figure}
\centering
\includegraphics[angle=0,width=12cm]{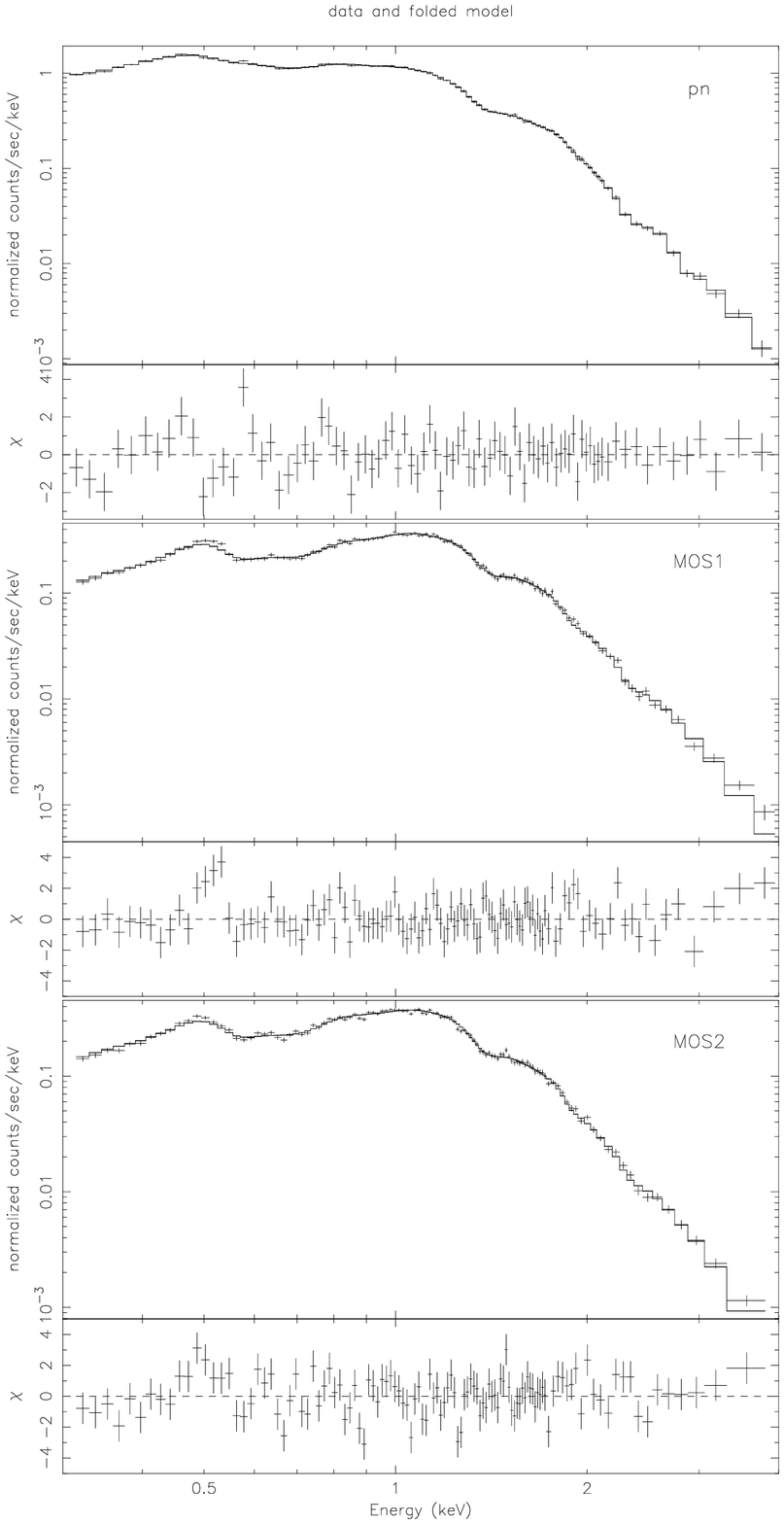}
  \caption{Fit of the phase-integrated data. The model (double blackbody plus line components)
  is described in the text.
From top to bottom, the panels show data from the pn, the MOS1 and the MOS2 cameras.
In each panel the data are compared to the model folded through
the instrumental response (upper plot); the lower plot shows the residuals in units of
sigma.
}
\label{spectra}
\end{figure}

Single component continuum models fail to reproduce the data (e.g.
$\chi^2_{\nu}>5$, 96 d.o.f. for a simple blackbody, including the line
components). The sum of a blackbody and a power law yielded a $\chi^2_{\nu}
\sim1.6$ (94 d.o.f., including the lines), predicting an excess of
counts at $E>$2.5 keV.
The best fit continuum curve
($\chi^2_{\nu}=1.15$, 94 d.o.f. including
the lines, see Fig.~\ref{spectra}, upper panel) is represented by the sum of two
blackbody functions. Assuming a distance of 2 kpc (Giacani et al. 2000), the cooler blackbody
(hereafter BB1) has a temperature
kT=0.163$\pm$0.003 keV and an emitting radius of 4.6$\pm$0.1 km
; the hotter (hereafter BB2) one has kT=0.319$\pm$0.001
keV and an emitting radius of 800$\pm$50 m.
The best value for the interstellar absorbing
column is N$_H$=(1.0$\pm$0.1)$\times 10^{21}$ cm$^{-2}$.

We note that a non-thermal, power law spectral component with
a luminosity similar to that observed for middle-aged pulsars
such as PSR B0656+14, Geminga and PSR B1055-52 (see e.g. Becker
\& Aschenbach 2002), namely a few $10^{30}$ erg s$^{-1}$ in the
0.3-4 keV range, would not be detectable at the distance of \ee.

The BB parameters differ from those reported in Paper II owing
to the updated calibrations used here. In particular, the flux at low
energy (in the 0.3-0.5 keV range) is found to be $\sim$30\% higher.

\begin{figure}[!htb]
\centering
\includegraphics[angle=-90,width=13cm]{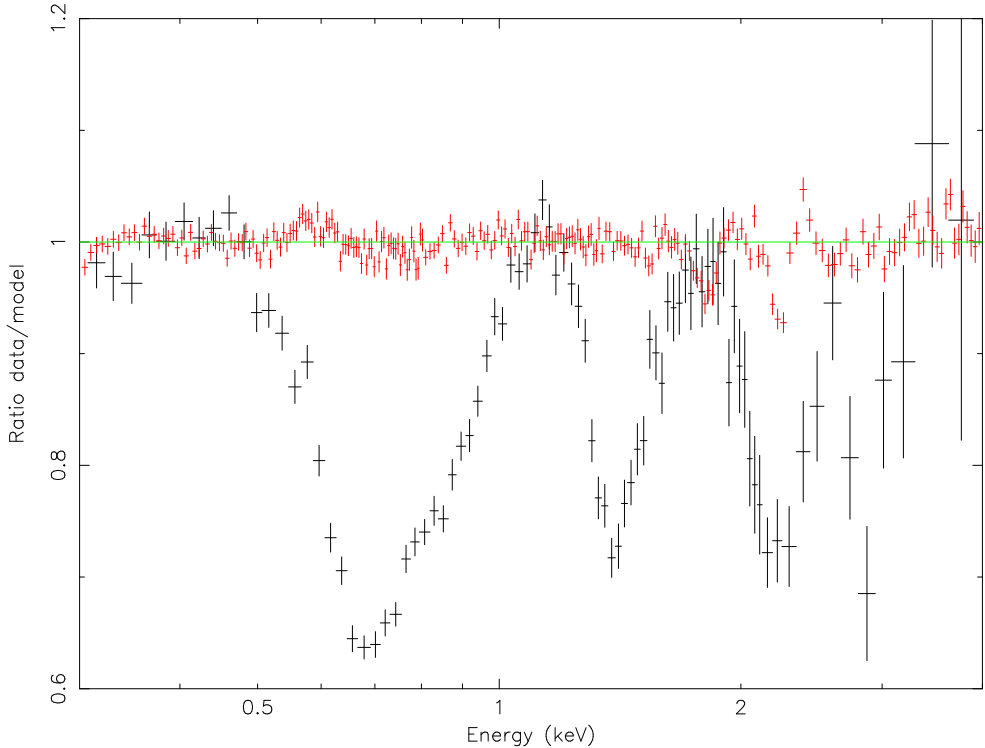}
  \caption{\label{3c273}
  Ratio of the data to the continuum best fit models for \ee (black points) and
for the bright quasar 3C\,273 (red points).
The EPIC/pn observation of 3C\,273 (performed on 2001/06/13; pn operated in
Small Window mode with the Medium filter) yielded $\sim$62 ksec
of good exposure time. PATTERN 0-4 events were extracted from a 43$''$ radius
region centered on the source position. Ad-hoc response files were generated.
The best fit ($\chi^2_\nu=1.29$, 1099 d.o.f.) spectral model in the range
0.3-6 keV is a slightly absorbed
(N$_H \sim 1.2 \times 10^{20}$ cm$^{-2}$) double power law
(photon indices $\alpha_{ph~1}\sim2.79$ and $\alpha_{ph~2}\sim1.43$;
 0.3-6 keV fluxes $F_1\sim4.47 \times 10^{-11}$ erg cm$^{-2}$ s$^{-1}$ and
 $F_2\sim8.67 \times 10^{-11}$ erg cm$^{-2}$ s$^{-1}$).
Deviations at the $\sim6-7$\% level are seen in the case of the
featureless 3C\,273 spectrum owing to residual miscalibrations at
the Si (1.839 keV) and Au (2.209 keV) edges. These artefacts can
be reproduced ($\chi^2_\nu=1.05$, 1093 d.o.f.) by two gaussian
lines in absorption centered at the edge energies, with
equivalent width of $\sim5.5$ eV and $\sim7.7$ eV for the Si and
the Au edge case, respectively. The $\sim$2.1 keV feature seen in
the spectrum of \ee (see text and Table~\ref{averes}) is markedly
stronger, with a $>$13 times higher equivalent width, and is
totally unrelated to the calibration leftovers which are never
seen in dim sources. }
\end{figure}

\begin{figure}
\centering
  \includegraphics[angle=-90,width=10cm]{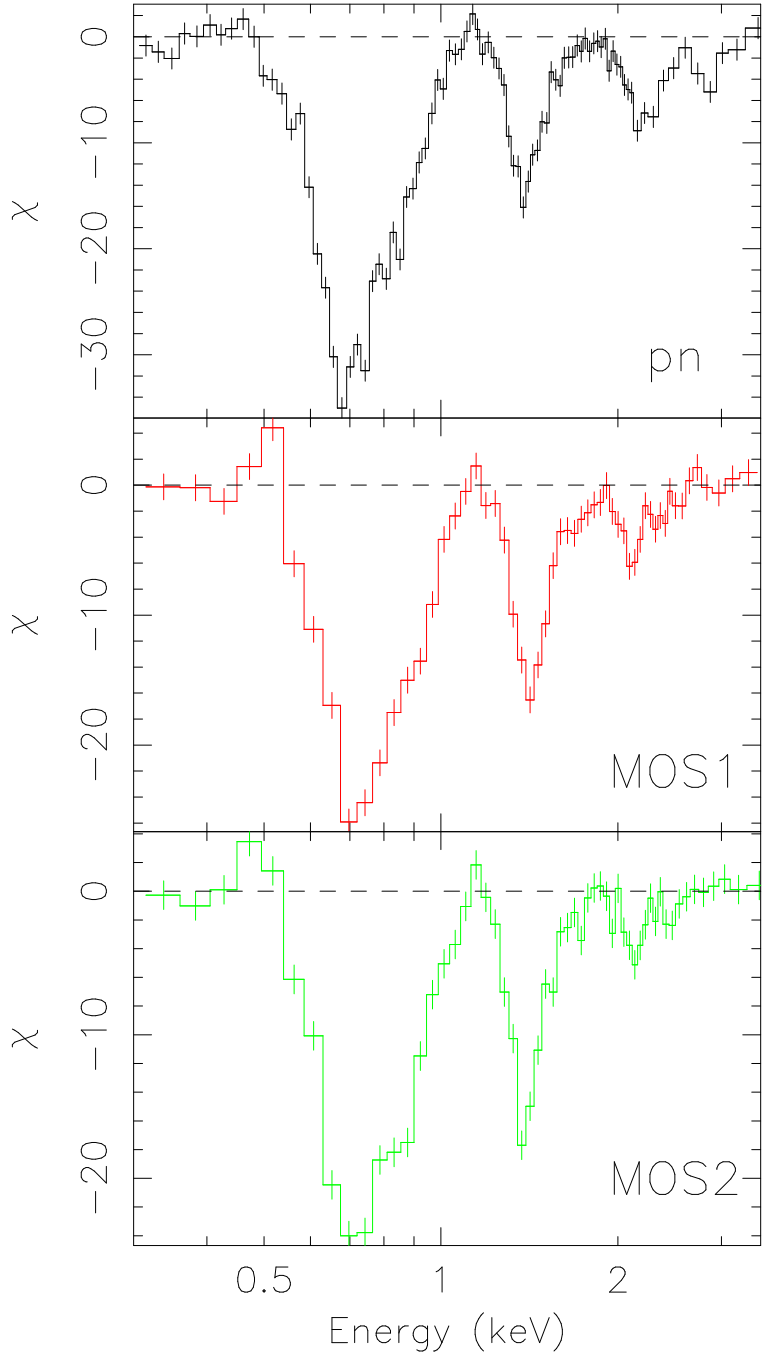}
  \caption{ Residuals in units of sigma obtained by comparing
  the data with the best fit thermal continuum model.
  The presence of four absorption features at $\sim0.7$ keV, $\sim1.4$ keV, $\sim2.1$ keV and $\sim2.8$ keV
  in the pn spectrum is evident. The three main features are also independently
  detected  by the MOS1 and MOS2 cameras. }

  \label{feat}
\end{figure}

Four absorption features are clearly seen in the
pn spectrum (see Fig~\ref{3c273} and top panel of Fig.~\ref{feat}) at the
harmonically spaced energies of
$\sim$0.7 keV, $\sim$1.4 keV, $\sim$2.1 keV and
$\sim$2.8 keV.

The different spectral continuum model
resulting from the improved
calibrations also yields  a more significant
detection of the
third and fourth features
with respect to Paper II.
 Using a simple gaussian in
absorption, we estimated with an F-test that the 2.1 keV
and the 2.8 keV features have a chance occurrence probability of
$\sim$10$^{-9}$ and  $\sim 10^{-3}$, respectively.

 In the spectral region encompassing the 2.1 keV absorption
feature, two instrumental edges, due to Si (1.839 keV) and Au
(2.209 keV), are present and, owing to minor miscalibrations in
the instrumental response, they may give rise to structured
residuals at a few \% level in high-statistic spectra (Kirsch
2003). To address the issue of the calibration accuracy in such
energy range, we have retrieved from the XMM-Newton Science
Archive the dataset relative to the observation of a very bright,
featureless source, the quasar 3C\,273. We plotted in
Fig.~\ref{3c273} (see caption for details on the data analysis)
the ratio of the data with respect to the best fit continuum model
for both 3C273 and \ee. In the case of 3C\,273, tiny deviations at
the $\sim$6-7\% level are seen at the expected energies of the
edges, with an Equivalent Width of $\leq$10 eV. As already pointed
out in Paper II, the case of \ee, is remarkably different. The
feature at $\sim$2.1 keV represents a 25-30\% depletion with
respect to the continuum level, while its Equivalent Width is
$\sim$100 eV. Thus, we rule out the possibility of an 
instrumental origin of the absorption feature near 2.1
keV in the spectrum of \ee.

The large number of  photons allows to determine with high
accuracy the profiles of the two main features. A simple gaussian
is inadequate to reproduce the broad dips, owing to their
asymmetric shape. The combination of two gaussian lines in
absorption can mimic their profile yielding a statistically good
fit ($\chi^2_{\nu} \sim 1.0$, 92 d.o.f.). We found that a good
fit can be also obtained using an asymmetric line profile having
the following analytic form:
\begin{equation}
A(E)=\exp[-a(E-E_0)^2-b(E-E_0)^3-c(E-E_0)^4]
\end{equation}
We note that this model does not have a direct physical meaning.
Rather, it is used to phenomenologically describe the data, since
it allows for an easy evaluation of the central energy of the
features and reproduces their shape with a small number of
parameters. We therefore adopt this model to fit the two main
features at 0.7 and 1.4 keV; the third and the fourth features,
owing to the lower statistics, can be well reproduced by simple
gaussian lines in absorption.

The results of the pn data analysis are reported in Table ~\ref{averes}.


We turn now to the MOS1 and MOS2 data, encompassing $\sim$73,800
and $\sim$75,600 photons in the 0.3-4 keV range. We found that,
also in these cases, the best fitting continuum model
($\chi^{2}_{\nu}$=1.19 for MOS1 - see Fig.~\ref{spectra}, middle panel,
$\chi^{2}_{\nu}$=1.47 for MOS2 due to the presence of several
small wiggles - see Fig.~\ref{spectra}, lower panel) is represented by the sum
of two blackbody curves. The temperatures and the emitting
regions are consistent with the results from the pn camera; the
interstellar absorption is found to be somewhat higher in the MOS
data (N$_{H}$=(1.3$\pm$0.1)$\times 10^{21} $cm$^{-2}$). We ascribe
this difference to the observed time evolution of the low energy
(E$<$0.5 keV) redistribution function of the MOS detectors
(Kirsch 2003). This effect is currently under investigation by
the calibration team; we therefore consider the pn measurement to
be more reliable.

Three absorption features are clearly detected in the MOS spectra
(see Fig.~\ref{feat}). The 2.1 keV feature has a chance
occurrence probability of order $10^{-3}$ per camera, as
estimated by means of an F-test. Owing to the lower statistics,
the 2.8 keV feature is only marginally detected by the MOS
cameras and was not included in the model. As in the case of the
pn, the shape of the two main features was found to be asymmetric
and well reproduced by the analytic profile described in Equation~1. The
parameters of the 0.7 and 1.4 keV features are consistent with
the pn results (see Table ~\ref{averes}.); however, due to the
smaller number of photons in the MOS data, some of the parameters
decribing the second feature are not well constrained. The lower
equivalent width of the 2.1 keV feature in the MOS should not be
a matter of concern since in this region of the spectrum the model
for the MOS and the pn (including a fourth broad feature at 2.8
keV) are different.



\subsubsection{Phase-resolved spectroscopy}

\begin{figure}
\centering
  \includegraphics[width=15cm,height=\textwidth,angle=-90]{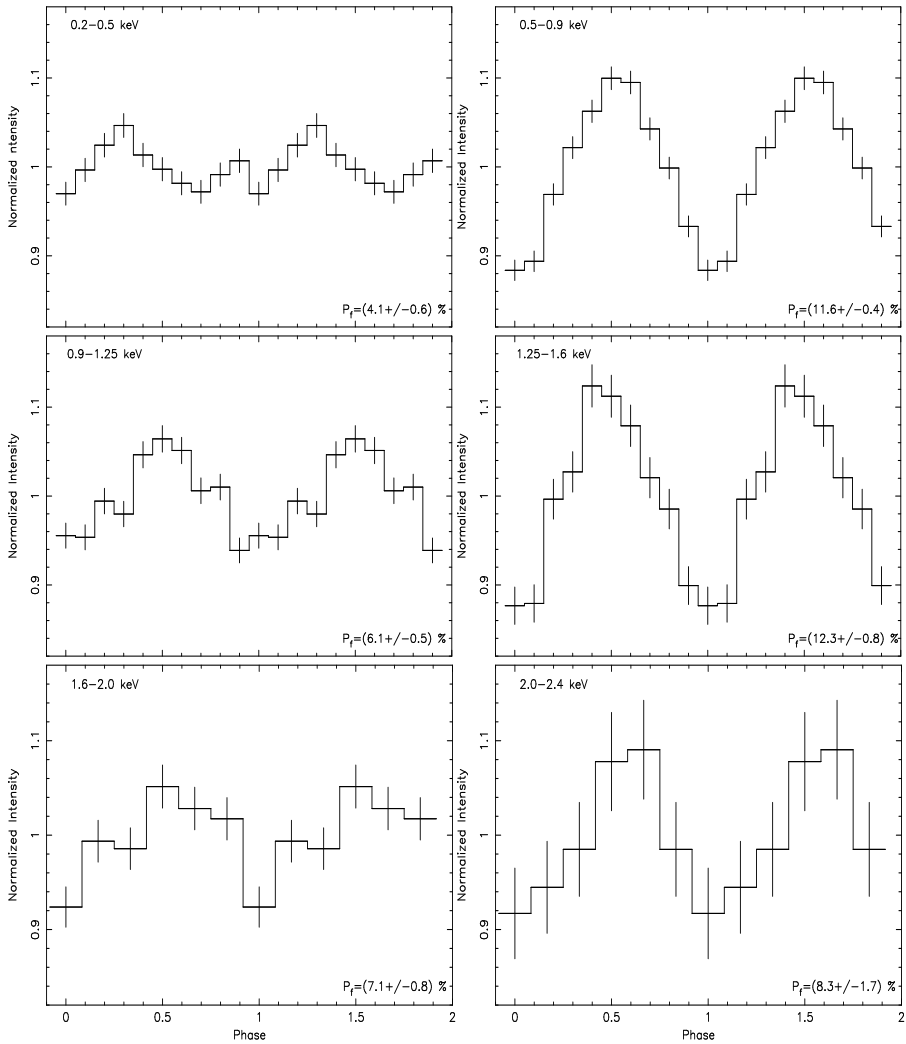}
  \caption{ Folded light curves of \ee. The energy ranges have been selected
  in order to put in evidence the impact of the phase variation of the features
  on the source pulsation.
  The three panels to the left show that in the spectral regions less affected by the main features
  the pulsed fraction is of order $\sim$4-6\%.
   The three panels to the right show that the pulsation is much higher ($\sim$10\%) in the
   energy ranges where the three main features are seen.
}
\label{varpf}
\end{figure}

As seen in our first EPIC observation of \ee, the absorption
features are phase-dependent (Paper I); in Paper II we showed
that the pulse phase variations of the spectrum are stronger in
correspondence of the features, while the continuum does not
change significantly. This is evident from Fig.~\ref{varpf}, where
the folded light curves have been plotted in six energy ranges
where lines are either dominant (right panels) or nearly absent
(left panels). The pulsed fraction is lower ($\sim$4-6\%) in the
spectral ranges where the features are less important, while is
definitely higher ($\sim$10\%) in correspondence to the three
main features.

In order to further investigate such an effect, we extracted
phase-resolved spectra.  Following Paper II, we selected the phase
intervals corresponding to the peak (phase interval 0.40-0.65
with respect to Fig.~\ref{lcs} and ~\ref{varpf}), the declining
part (phase 0.65-0.90), the minimum (phase 0.90-1.15) and the
rising part (phase 0.15-0.40) of the folded light curve.

The resulting spectra were fitted allowing both the thermal
continuum and the lines to vary. For the sake of simplicity, the
fourth feature, owing to its lower significance, was not included
in the model.

\begin{figure}
\centering
  \includegraphics[angle=-90,width=9cm]{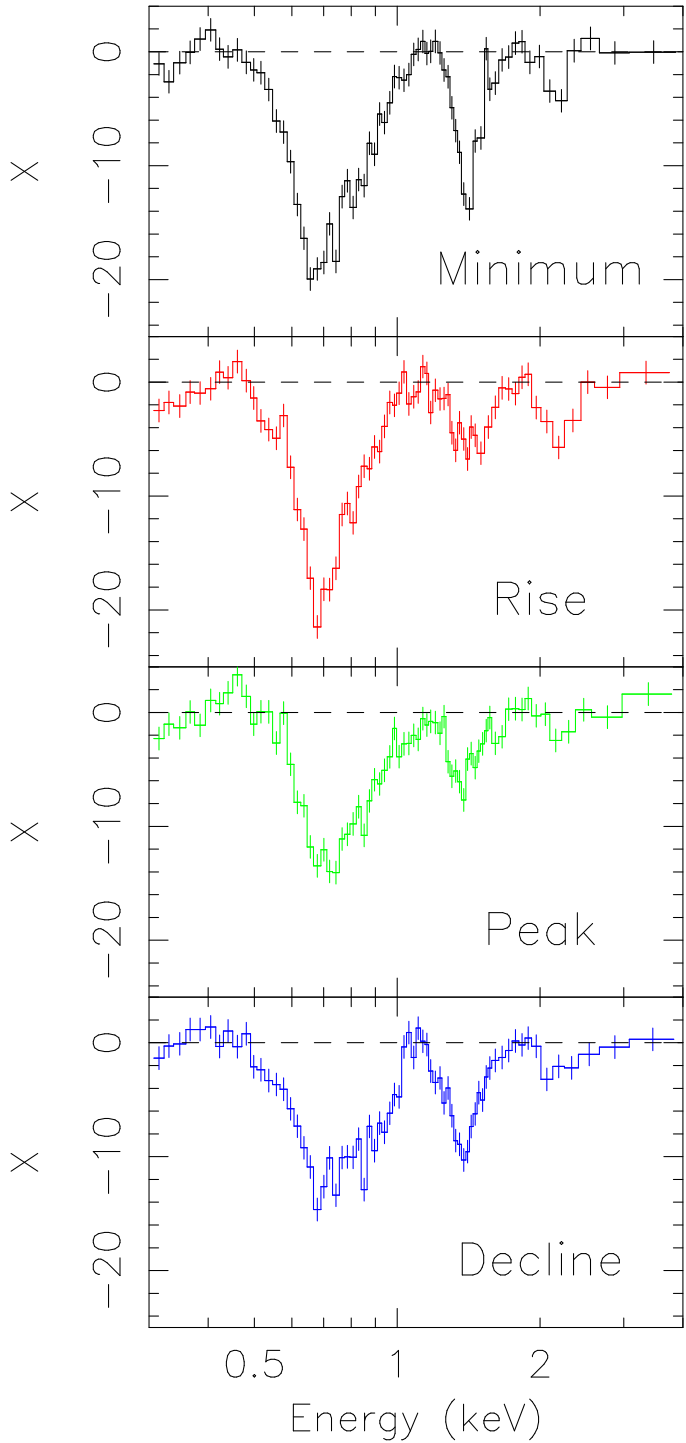}
  \caption{ Residuals in units of sigma obtained by comparing
  the data with the best fit continuum model (line components removed) for the
  phase-resolved spectra.
  The variations in width, depth and shape of the absorption features as
  a function of the pulse phase are evident.
  Structured residuals are observed in the range 0.4-0.7 keV for the
``rise'' spectrum and (more significantly) in the range 0.9-1.2 keV in the ``decline'', where broad emission
features could be present. We did not attempt to model these structures, which are
likely artefacts due to an inadequate description of the main absorption features' profile.
}
\label{phaseresfeat}
\end{figure}

We give the best fit parameters, describing the phase resolved
spectra, in Table ~\ref{phaseresres}. Note that systematic
uncertainties were not included, as stated in Sect.~\ref{spec}.
In Fig.~\ref{phaseresfeat} we plot the residuals with respect to
the continuum to show the phase variations of the features.

The results of the phase resolved spectroscopy can be summarized as follows:
\begin{itemize}
\item The two components of the continuum vary slightly (see Table~\ref{phaseresres})
both in temperature and in flux with the pulse phase, accounting for $\sim$3-5\%
of the source pulsation. A comparison with
the values reported in Sect.~\ref{timing} and an inspection of Fig.~\ref{varpf}
clearly shows that the phase variation of
the features is largely responsible for the observed pulsation of the source.

\item the features are strongly dependent on the pulse phase; variations in
their width, depth (see Table~\ref{phaseresres}) and shape (see
Fig.~\ref{phaseresfeat}) are evident. The central energy of the
0.7 keV feature (as derived from the phenomenological model) vary
at most of 6\%, while displacements of the other features are not
significant. The 2.8 keV feature is marginally detected only
during the minimum and the rise intervals.
\item The relative intensity of the first three features varies in phase.

\end{itemize}

\section{Optical data analysis}
The field of 1E 1207$-$5209 was observed in Service Mode between April
and July 2002 with the 8.2-meter  UT-1 Telescope (Antu) of the ESO VLT
(Paranal Observatory). Observations were  performed with the the FOcal
Reducer and Spectrograph 1 (FORS1) instrument,
operated as imager in its
standard  resolution  mode,  with  a  pixel size  of  0\farcs2  and  a
corresponding  field  of view  of  $6\farcm8  \times 6\farcm8$,  large
enough  to  include   many   reference  stars  for  a  precise  CCD
astrometry.    Images   were   acquired   through   the   Bessel   $V$
 and $R$ filters  for  a  total
integration  time of  $\sim $2  and  3 hrs,  respectively.  The  total
integration  time was  split  in short  exposures  of 100  s to  avoid
saturation of two  relatively bright stars (star A  and B - Bignami
et al.   1992) close to  the position of  our target and allow  for an
efficient cosmic ray filtering.  Table ~\ref{optjournal} reports the complete journal
of  the  observations.   Exposures   were  acquired  under  dark  time
conditions at airmass $\langle\sec z\rangle \sim 1.5$ and with average
seeing  of $\sim  0\farcs7$ and  $\sim 0\farcs9$  in the  $V$  and $R$
bands,  respectively.  Atmospheric conditions  were good,  with little
wind and humidity always below 10\%.  All nights were photometric.

Standard  reduction steps including  debiassing and  flatfielding were
  applied through  the ESO  FORS1 data  reduction  pipeline.  Flux
calibration was  performed using images of  photometric standards from
the  Landolt fields  (Landolt 1992),  yielding extinction\footnote{The
atmospheric extinction  correction was computed according  to the most
recent    coefficients   measured    for    Paranal,   available    at
http://www.eso.org/observing/dfo/quality/FORS1/qc/photcoeff/}       and
color-corrected  zero-points  with  an   overall  accuracy  of  a  few
hundredths  of magnitude.  Night-to-night  zero-point variations  were
found to be  below 0.04 magnitudes.  For each  filter, images taken in
different nights have been registered  and averaged using a 3 $\sigma$
clipping algorithm to reject cosmic ray hits.

In  order to  register accurately  the  target position  on the
FORS1 images, we have  recomputed the image astrometry using  as
a reference the positions  of 20 well-suited,  (i.e. not
extended, not  too faint, and not too close to the CCD edges)
stars selected from the Guide Star Catalogue II. The overall
accuracy of  our astrometric solution  was of $0\farcs17$ per
coordinate.

\begin{figure}
\centering
  \includegraphics[angle=0,width=\textwidth]{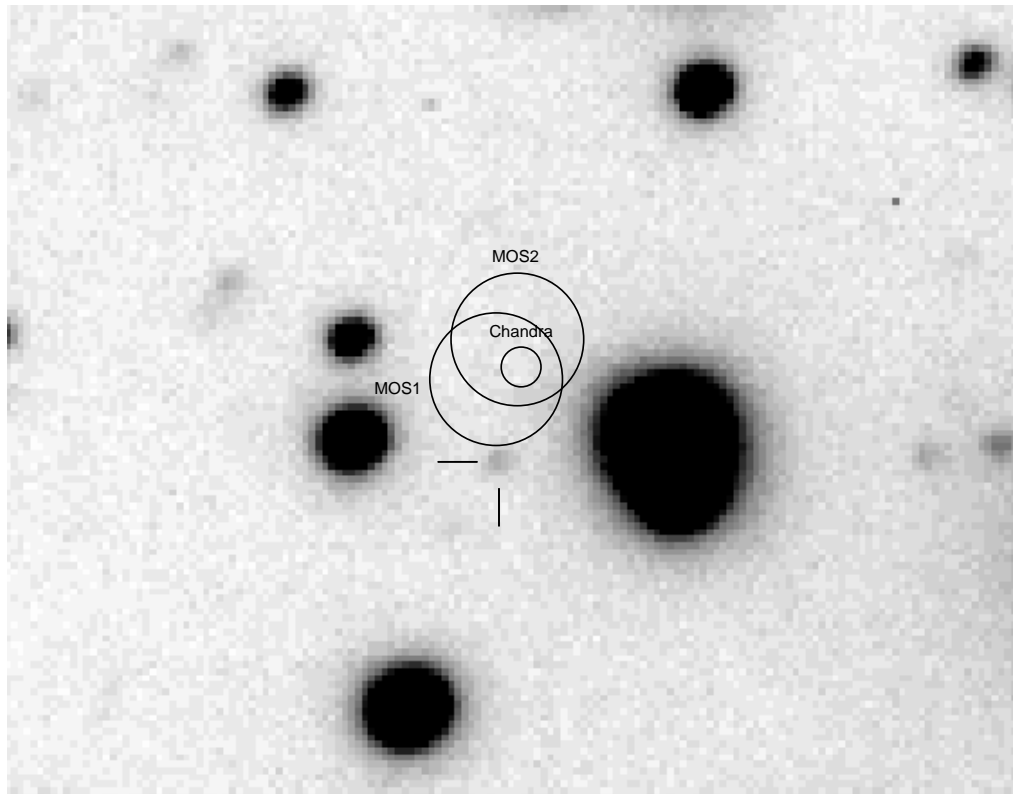}
  \caption{Combined $V$-band FORS1 image of the central region of the
  1E 1207$-$5209 field.
  The circles indicate the expected source positions computed from the
  Chandra/ACIS  (0.6$''$ radius),
  MOS1 and MOS2 (2$''$ radius) data.
  The good agreement among the three independent positions gives us confidence
  on the overall correctness of our analysis.
  No sources down to R$\sim$27.1 and  V$\sim$27.3 are seen inside the
  error circles.
The object labelled with the two small ticks,
  visible also in the NTT observation of Bignami et al. (1992),
  falls just outside of the MOS1 error circle.  Its
time-integrated magnitude in the combined  images is $V \sim
25.8$ and $R \sim  24.5$, with  uncertainties of $\sim  0.1$
magnitudes. Its profile is  consistent with a stellar-like
object. Its flux varied of $\sim 0.4$ magnitudes in the $V$-band
images (from $V \sim 26.2$ to $V \sim 25.7$), collected during 3
nights, and of $\sim 1.5$ magnitudes in the $R$-band images,
collected during 4 nights. It is most probably a background AGN. }
\label{vltfield}
\end{figure}

Fig.~\ref{vltfield} shows the inner portion  of the combined FORS1 $V$-band
image centered
on the  target position, with the MOS1, MOS2 and ACIS error circles
superimposed.

A faint object (marked with the  two ticks in
Fig.~\ref{vltfield}) is detected just outside the southern edge
of the MOS1 error circle and showed variability along the time
span covered by our observations (see caption of
Fig.~\ref{vltfield}). Its position, in any case, falls  more than
2$''$  away from the most probable one and we can rule it out as a
potential counterpart of \ee.

No
candidate counterpart  is detected  in the
Chandra error circle (nor in the intersection of the MOS ones)
down to R$\sim$27.1 and  V$\sim$27.3, which we assume as upper
limits on the optical flux of 1E 1207$-$5209.

\begin{table*}[h]
\begin{center}
\begin{tabular}{lccccc} \hline \hline
Date & Filter &   No.\ of exp. & Exposure(s) & seeing (") & airmass \\ \hline
2002 April 22    & $R$ & 43 & 4\,300 & 1.0"  & 1.33 \\ \hline
2002 May 10      & $R$ & 4  & 400    & 0.8"  & 1.034\\  \hline
2002 July 9      & $R$ & 38 & 3\,800 & 0.58" & 1.134 \\
2002 July 9      & $V$ & 26 & 2\,600 & 0.58" & 1.134 \\\hline
2002 July 13     & $V$ & 26 & 2\,600 & 0.52" & 1.054\\ \hline
2002 July 14     & $V$ & 26 & 2\,600 & 0.96" & 1.017\\ \hline
2002 July 15     & $R$ & 26 & 2\,600 & 0.96" & 1.017\\ \hline \hline
\end{tabular}
\end{center}
\caption{\label{optjournal} Summary   of the  optical observations  of   the field of 1E
1207$-$5209 performed with the FORS1 instrument  at VLT/Antu.
}

\end{table*}

\section{Discussion}

\subsection{The thermal continuum emission}

The X-ray spectral energy distribution of \ee shows
a continuum emission of purely thermal origin
which is well reproduced by the sum of two blackbody curves.

Thanks to the unprecedented throughput of the EPIC pn camera, we
detected with high statistical significance the variation of the
X-ray continuum with the pulse phase. The two blackbody
components show a low amplitude ($\leq$5\%) modulation in both
temperature and  flux. This strongly suggests a non-uniform
temperature distribution on the neutron star surface. The colder
blackbody component (kT$\sim$0.16 keV) is emitted from a quite
large fraction of the surface ($R\sim4.5$ km), while the hotter
(kT$\sim$0.3 keV) is possibly coming from a heated polar cap
($R\sim800$ m). The use of a more physical continuum model (i.e.
a magnetized atmosphere model) could yield lower surface
temperatures and larger emitting radii, as observed in several
cases (e.g. for Vela, Pavlov et al. 2001). The phase shift of
$\sim 90^{\circ}$ observed between the peaks of the light curves
below and above 0.5 keV requires an energy-dependent asymmetry in
the emission pattern.

The optical upper limits fall more than two orders of magnitude
above the extrapolation of the best fit X-ray blackbody curves.
For a distance of  2 kpc, after dereddening  for an absorption
$A_{V} \sim$0.6, the upper limit on the optical  luminosity
$L_{opt}$ is $\sim  3.4 \times 10^{28}$  erg s$^{-1}$.  Using the
estimate  of the total rotational energy loss obtained from the
X-ray timing, $\dot{E}_{rot}$ = (7.4$\pm$1.5)$\times$10$^{33}$
erg s$^{-1}$, we  derive  an optical emission efficiency
$L_{opt}/\dot  E_{rot} < 4.6 \times 10 ^{-6}$. This value is
comparable to the optical emission efficiency of middle-aged
neutron stars like PSR B0656+14 and Geminga.

\subsection{The absorption features as cyclotron absorption lines}
\label{cyclo} The reanalysis of the long EPIC observation, taking
advantage of the recent improvements in the instrument
calibration, yielded a more significant   detection of the
absorption features at 2.1 keV and 2.8 keV.  The conclusion of
Paper II is therefore strenghtened: the presence of four
absorption features, having central energies very close to the
ratio 1:2:3:4, coupled with their observed phase variation, is
naturally explained by cyclotron absorption from one fundamental
and three harmonics.

The absorption features are seen during all
phase intervals.
The absorbing layer is therefore surrounding most (all) of the X-ray emitting region.
The observed $\Delta E/E$ of the features, about 0.1$\div$0.3 depending
on the phase interval, implies that the variation of the
magnetic field across the line forming region ($\Delta E/E \sim \Delta B/B$) is
very small, constraining both the thickness and the latitudinal extent of the
absorbing layer.
Assuming for simplicity a standard dipole configuration,
the  radial dependence of the magnetic field
yields  $\Delta B/B \sim 3 \Delta r/r$. The data therefore
constrain the thickness
$\Delta r$ to (0.3-1)r$_{10}$ km, depending on the phase
interval, where r$_{10}$ is the distance (in units of 10 km) of the
absorbing layer from the neutron star centre.
Moreover, considering the dependence of the dipole field on the latitude,
B$\sim (1+3\,\cos^2 \theta)^{\frac{1}{2}}$, the measured
line width implies that the  absorbing region is limited
in latitude to an interval of 15-30 degrees from the equator, or 30-50 degrees from
the pole, depending on the phase interval.

\subsection{The cyclotron magnetic field and the pulsar slow-down}

The cyclotron interpretation of the absorption features allows for a direct
measure of the neutron star magnetic field.
This can be compared with the magnetic field value estimated from the observed timing
parameters, B = (2.6$\pm$0.3)$\times$10$^{12}$ G, assuming an uniform slow-down due to magneto-dipole braking.

Assuming that the 0.7 keV feature corresponds
to the fundamental cyclotron energy for electrons or for protons,
the inferred magnetic field would be of 0.6(1+z)$\times10^{11}$ G or
1.2(1+z)$\times10^{14}$ G, respectively; z represents the gravitational
redshift where the absorption occurs.
Two hypotheses about the absorbing region position can be explored: (i)
in the atmosphere, close to the NS surface and (ii) in the magnetosphere, at a few
stellar radii.

In the case of electrons, if the cyclotron features are formed
close to the surface, the fundamental cyclotron energy of
$\sim$0.7 keV yields B$\sim$8$\times 10^{10}$ G , assuming a
standard 25\% gravitational redshift. This value is $\sim$ 30
times lower than expected from the observed P and $\pdot$. A
``braking problem'' arises: some additional torque should be
acting in order to produce the observed spin-down of the neutron
star. A  disk from the fallback of supernova ejecta could induce
such an additional braking if the system is in the propeller
regime. Debris disks have been invoked (Chatterjee et al. 2000)
to account for the properties of anomalous X-ray pulsars (AXPs,
Mereghetti et al. 2002b).  The disk models predict an excess in
the optical emission (e.g. Perna et al. 2000); indeed, the
detection of a few AXPs in the IR band (e.g. Israel et al. 2003
and references therein) recently renewed attention to the disk
hypothesis. Although the actual properties of such systems are
largely unconstrained, in the case of \ee the optical upper
limits would require any hypothetical disk to be
underluminous\footnote{a rough estimate with a standard, purely
dissipative disk model (assuming an inner radius equal to the
magnetospheric radius, $r_M \sim 1.6 \times 10^8$ cm and a rate
of material reaching $r_M$ of $\sim6\times10^{-11}$ M$_{\odot}$
y$^{-1}$) yields a luminosity $\sim$400 times higher than allowed
by the VLT observations}. As a consequence, this scenario appears
rather unlikely. Alternatively, the cyclotron absorption could
take place in the magnetosphere, as discussed by Sanwal et al.
(2002). Sturner \& Dermer (1991) and Dermer \& Sturner (1994)
have suggested that a layer of accreted plasma could form in the
magnetosphere of a neutron star, at a height of a few stellar
radii, supported by radiation pressure. An alternative
possibility is represented by the presence of a ``blanket'' of
$e^{+}$/$e^{-}$ pairs formed in the closed line region of the
neutron star magnetosphere at an altitude of $\sim$3 stellar
radii (Wang et al. 1998). Although the actual suspended mass, the
density, as well as the stability of the suspended absorbing
layer are highly uncertain, we note that in this picture the
inferred surface magnetic field would be very close to the value
expected from the spin parameters.

In the case of proton
 absorption, the inferred magnetic field would be at least of $1.6\times10^{14}$ G (absorption
occurring at the surface).
Such a ``magnetar''-like field would not be free from problems when considering the
measured $\pdot$.
The spin-down expected from the standard dipole formula would be much larger than observed,
 $ \sim 6\times10^{-11}$ s s$^{-1}$.
A possibility for solving this problem could be the presence of higher multipole components
in the magnetic field of the neutron star, as suggested
by Sanwal et al. (2002).
The surface value could exceed $10^{14}$ G, while the
braking would be driven by the dipole component, dominating at large radii.

\subsection{A $\dot{P}$ problem ?}

The magnetic field inferred from the observed spin-down
is hardly conciliable with the independent estimates offered by the cyclotron interpretation
of the absorption features.
Furthermore, our refined $\dot{P}$ value (see Sect.~\ref{timing}) confirms
the ``age problem'' (Pavlov et al. 2002):
the characteristic age of \ee, $\tau_c$ = (4.7$\pm$1.0)$\times$10$^{5}$ yrs,
is more than 50 times higher than the age of the associated supernova remnant ($\tau_{SNR}$ $\sim 7$ kyrs).

To reconcile this discrepancy the possibility of a birth period
close to the present one was proposed (Pavlov et al. 2002; Paper
I). The same born-slow hypothesis was suggested to solve similar
age inconsistencies in few other cases (e.g. PSR J1811-1925 in G11.2-0.3, Kaspi et al. 2001; 
PSR J0538+2817 in S147, Kramer et al. 2003). Such sources represent a big problem
for massive stars core-collapse theory, since it is difficult to
explain initial spin periods as large as a few tens of
milliseconds (Heger et al. 2003).

Of course, all of the above assumes \ee to be a smoothly slowing
down pulsar. If the period evolution of the source is not
monotonic, our measurement of the period derivative, based on a
set of sparse observations, could be wrong. For example, glitches
could dominate the long-term spin-down of the source, which would
appear lower than that due to the magnetodipole radiation. The
$\dot{P}$ measurement might also be affected by Doppler shift, if
the neutron star is in a binary system. In this case, the optical
upper limits ($R>27.1$, $V>27.3$), for a distance of 2 kpc and an
absorption $A_V\sim0.6$, exclude any main sequence star as a
possible companion to \ee, leaving open only the possibility of
 a degenerate object. 

 We note that a very recent (June 2003) Chandra observation
of \ee did not help to clarify the issue of the period evolution of the source
(see Zavlin et al. 2003).

\subsection{\ee: out of the chorus line ?}
An important problem still lacking a solution concerns the uniqueness of
the phenomenology of \ee in the isolated neutron stars panorama.
High quality X-ray observations
have been performed on several isolated neutron stars of various kind.
The detection of absorption features has been claimed only in
the Soft Gamma Repeater 1806-20 (Ibrahim et al. 2002,2003),
in the Anomalous X-ray Pulsar 1RXS 170849--400910 (Rea et al. 2003)
and in the isolated neutron stars RBS 1223 (Haberl et al. 2003)
 and RX J1605.3+3249 (van Kerkwijk 2003).
None of these cases is comparable to the phenomenology of \ee.

A possible explanation invokes selection effects. \ee could be an
``unconventional'',
low magnetic field neutron star ($< 10^{11}$ G as in the electron, near-surface cyclotron
scenario). In sources with more typical magnetic fields of order
$10^{12}$ G the cyclotron energy
would lie in the range of the tens of keV, where the surface thermal
emission is very low and where sensitive X-ray observations have not yet been obtained.
If the magnetic field of \ee, on the contrary, is closer to the typical
value of $10^{12}$ G (as in the magnetospheric cyclotron scenario),
the absence of cyclotron features in other sources could be
explained by a limited lifetime for the  absorbing layer.

An answer to the ``uniqueness problem'' could come
only through a better understanding of
the overall properties of the Central Compact Obiects in supernova remnants,
 the fraternity of \ee. These sources are
supposed to be the youngest members of the radio-quiet neutron stars family
(including also Anomalous X-ray Pulsars, Soft Gamma Repeaters and Dim Thermal
Neutron Stars), but their physics remain elusive.
We do not understand the lack of radio emission, the lack of X-ray
pulsations (\ee being unique, at the moment, also in this aspect of the phenomenology).
The fallback of supernova ejecta
is possibly playing an important role, driving their multiwavelength emission,
their spin-down and their evolution,
 as suggested by Alpar (2001). Deeper observations
of these sources could shed light to the overall scenario. Only in this
perspective we will have an answer to the question whether \ee is indeed
an unique object,
or is simply in a standard (transient) phase of the evolution of a young
neutron star.

\clearpage

\begin{table}[h]
\caption{\label{averes}{\it  Results of phase-integrated spectroscopy }}


\begin{center}

\begin{tabular}{lccc}

\hline

\smallskip

          & pn &   MOS1  & MOS2 \\


\hline

\smallskip

& & & \\

N$_H$ (10$^{21}$ cm$^{-2}$) & 1.0$\pm0.1$  & 1.4$\pm0.1$       & 1.3$\pm0.1$   \\
\smallskip
kT$_{BB1}$   (keV)                &  0.164$\pm$0.001       & 0.165$\pm$0.002    & 0.168$\pm$0.002 \\
\smallskip
R$_{BB1}^{(a)}$  (km)        &  4.5$\pm$0.1         &  4.6$\pm$0.1         & 4.6$\pm$0.1   \\
\smallskip
kT$_{BB2}$   (keV)                &  0.319$\pm$0.002       & 0.322$\pm$0.002    &0.320$\pm$0.002 \\
\smallskip
R$_{BB2}^{(a)}$  (km)        &  0.83$\pm$0.02         &  0.83$\pm$0.03         & 0.83$\pm$0.03   \\
\smallskip
\\
\smallskip
E$_{1}$  (keV)              &  0.68$\pm$0.01       & 0.69$\pm0.01$ &0.69$\pm0.01$ \\
\smallskip
a$_{1}$           &  77$^{+2}_{-3}$       & 80$\pm$3        & 82$\pm$3       \\
\smallskip
b$_{1}$           &  -368$^{+7}_{-22}$       & -419$\pm$15        & -412$\pm$15       \\
\smallskip
c$_{1}$           &  616$^{+26}_{-44}$      & 749$\pm$40        & 684$\pm$40       \\
\smallskip
EW$_{1}$      (eV)          & 99$^{+4}_{-7}$   & 105$^{+6}_{-7}$  & 104$^{+6}_{-7}$    \\
\smallskip
FWHM$_{1}$      (keV)          & 0.24$\pm0.01$   & 0.24$\pm0.02$  &   0.24$\pm0.02$  \\
\smallskip
E$_{2}$  (keV)              &  1.36$\pm$0.01       & 1.39$\pm0.02$ & 1.37$\pm0.02$ \\
\smallskip
a$_{2}$           &    96$^{+8}_{-32}$     & 89$^{+33}_{-11}$        & 96$^{+27}_{-8}$       \\
\smallskip
b$_{2}$           &  -370$^{+50}_{-20}$       & -360$^{+250}_{-30}$        & -220$^{+270}_{-35}$       \\
\smallskip
c$_{2}$           &  488$^{+145}_{-55}$       &   220$^{+1000}_{-55}$      &   500$^{+1000}_{-80}$     \\
\smallskip
EW$_{2}$      (eV)          & 66$\pm6$   & 64$\pm7$  & 67$\pm7$    \\
\smallskip
FWHM$_{2}$      (keV)          & 0.18$\pm0.02$   & 0.19$\pm0.03$  &   0.20$\pm0.03$  \\
\smallskip
E$_{3}$     (keV)           &  2.14$\pm$0.03       & 2.12$\pm$0.07        & 2.12$\pm$0.08       \\
\smallskip
$\sigma_{3}$  (keV)         &  0.17$\pm$0.03       &  0.13$^{+0.10}_{-0.04}$       &   0.10$^{+0.10}_{-0.04}$    \\
\smallskip
$EW_{3}$     (eV)           & 94$\pm15$  & 48$\pm20$   & 45$\pm20$    \\
\smallskip
E$_{4}$     (keV)           &  2.83$\pm$0.05       & --        & --       \\
\smallskip
$\sigma_{4}$  (keV)         &  0.11$^{+0.10}_{-0.04}$       &  --       & --      \\
\smallskip
$EW_{4}$     (eV)           & 81$^{+20}_{-30}$  & --   & --    \\
\smallskip
F$_{0.3-4 keV}^{(b)}$ (erg cm$^{-2}$ s$^{-1}$)  &  2.24$\times10^{-12}$ & 2.10$\times10^{-12}$   &   2.17$\times10^{-12}$     \\
\smallskip
\smallskip
L$_{X}^{(c)}$ (erg s$^{-1}$) &   2.1$\times10^{33}$ &   2.2$\times10^{33}$ & 2.2$\times10^{33}$  \\
\smallskip
$\chi^{2}$/dof              &    1.15            & 1.19               &    1.47\\
\smallskip
dof                         &   94               & 154                 & 151  \\

\hline
\end{tabular}


\end{center}

$^a$ Radius at infinity for an assumed distance of 2 kpc.

$^b$ Observed flux.

$^c$ Bolometric luminosity for d=2 kpc.

All the errors are at the 90\% c.l. for a single interesting parameter
\end{table}

\begin{table}[h]
\caption{\label{phaseresres}{\it  Results of phase resolved spectroscopy }}


\begin{center}

\begin{tabular}{lcccc }

\hline

\smallskip
          &    Minimum  &   Rise &  Peak &   Decline \\
\hline
\smallskip
& & & \\
N$_H$ (10$^{21}$ cm$^{-2}$) fixed & 1.0   & 1.0        &  1.0   & 1.0    \\
\smallskip
kT$_{BB1}$     (keV)    &  0.157$\pm$0.001 &  0.156$\pm$0.001   &
0.171$\pm$0.002  & 0.164$\pm$0.002   \\
\smallskip
R$_{BB1}^{(a)}$  (km)   &  4.67$\pm$0.03       & 4.85$\pm$0.06   &
4.01$\pm$0.03
& 4.20$\pm$0.04    \\
\smallskip
kT$_{BB2}$     (keV)    &  0.297$\pm$0.002 &  0.302$\pm$0.002   &
0.293$\pm$0.002  & 0.290$\pm$0.002   \\
\smallskip
R$_{BB2}^{(a)}$  (km)   &  1.00$\pm$0.01       & 0.98$\pm$0.03    &
1.04$\pm$0.01    & 1.12$\pm$0.03    \\
\smallskip
E$_{1}$  (keV) &  0.664$\pm$0.005 &  0.679$\pm$0.005    & 0.703$\pm$0.006 &
0.668$\pm$0.012\\
\smallskip
a$_{1}$   &  94.4$\pm$1.5   &  98.2$^{+3.4}_{-2.6}$    &
78.7$^{+1.9}_{-1.8}$ &
70.1$^{+1.5}_{-12.1}$  \\
\smallskip
b$_{1}$   &  -499$^{+6}_{-5}$   &  -464$^{+15}_{-9}$    & -320$^{+7}_{-6}$ &
  -
403$^{+11}_{-9}$  \\
\smallskip
c$_{1}$   &  848$^{+21}_{-20}$ & 762$^{+51}_{-34}$  &   424$^{+23}_{-17}$   &
724$^{+21}_{-37}$      \\
\smallskip
\smallskip
EW$_{1}$  (eV)  &   114$^{+4}_{-5}$ & 103$^{+5}_{-6}$ &   92$^{+5}_{-5}$   &
93$^{+5}_{-3}$    \\
\smallskip
FWHM$_{1}$ (keV)  &  0.22$\pm$0.01   &  0.19$\pm$0.01    & 0.21$\pm$0.01 &
0.32$\pm$0.02  \\
\smallskip
r$_{1}^{(b)}$  &   0.46$\pm$0.02   &   0.43$\pm$0.02   &     0.29$\pm$0.02
&
0.30$\pm$0.02    \\
\smallskip
E$_{2}$  (keV) &  1.37$\pm$0.01 &  1.38$\pm$0.02    & 1.33$\pm$0.01 &
1.35$\pm$0.01\\
\smallskip
a$_{2}$   &  147$\pm$11   &  34.5$\pm$0.5    & 171$^{+16}_{-5}$ &
63.3$^{+0.8}_{-0.5}$  \\
\smallskip
b$_{2}$   &  -690$^{+62}_{-38}$   &    -7.33$^{+0.09}_{-0.01}$  &
-981$^{+73}_{-
19}$ &  -121$^{+1}_{-2}$  \\
\smallskip
c$_{2}$   &  1087$^{+333}_{-137}$   &   0.213$^{+0.010}_{-0.002}$   &
1629$^{+216}_{-68}$ &   60.3$^{+1.0}_{-0.8}$ \\
\smallskip
EW$_{2}$ (eV)  &   82$\pm$6  & 66$\pm$8  &  54$\pm$6 &  93$\pm$7 \\
\smallskip
FWHM$_{2}$  (keV) &  0.14$\pm$0.02   &  0.35$\pm$0.04    & 0.14$\pm$0.02 &
0.21$\pm$0.02  \\
\smallskip
r$_{2}^{(b)}$  &   0.41$\pm$0.02   &   0.19$\pm$0.03   &     0.19$\pm$0.03
&
0.32$\pm$0.02    \\
\smallskip
E$_{3}$  (keV) &  2.16$\pm$0.05 &   2.15$\pm$0.06   & 2.12$\pm$0.04 &
2.08$\pm$0.06\\
\smallskip
$\sigma_{3}$  (keV) &  0.09$\pm$0.04 &  0.14$\pm$0.05    & 0.08$\pm$0.06 &
0.08$\pm$0.06\\
\smallskip
EW$_{3}$  (eV) &  64$^{+19}_{-19}$ &   102$^{+35}_{-40}$   &
28$^{+13}_{-13}$ &
17$^{+11}_{-11}$\\
r$_{3}^{(b)}$  &  0.25$\pm$0.06 & 0.29$\pm$0.06 & 0.13$\pm$0.06  & 0.15$\pm$0.06 \\
\smallskip
$\chi^{2}$/dof             &  1.09 &    1.20            & 1.35
&
1.01 \\
\smallskip
dof                         & 79   &   79              & 78
& 78
\\

\hline

\end{tabular}

\end{center}

$^a$ Radius at infinity for an assumed distance of 2 kpc.

$^b$ Relative line depth =  1. -- F (E$_{line}$) / F$_{C}$ (E$_{line}$),
where F is the total flux and F$_{C}$ is the flux of the continuum only

All the errors are at the 90\% confidence level for a single interesting
parameter

\end{table}

\end{document}